\documentstyle[aps,graphicx]{revtex}

\begin{document}

\twocolumn[\hsize\textwidth\columnwidth\hsize\csname @twocolumnfalse\endcsname

\title{Canonical and Microcanonical Ensemble Approaches to
Bose-Einstein Condensation:
The Thermodynamics of Particles in Harmonic Traps}
\author{K.~C. Chase, A.~Z. Mekjian and L. Zamick}
\address{Rutgers University, Department of Physics, Piscataway, NJ 08855-0849}
\date{\today}

\maketitle

\begin{abstract}
The thermodynamic properties of bosons moving in a harmonic trap
in an arbitrary number of dimensions are investigated in the grand
canonical, canonical and microcanonical
ensembles by applying combinatorial techniques developed earlier
in statistical nuclear fragmentation models.  Thermodynamic functions
such as the energy and specific heat are computed exactly in
these ensembles.  The occupation of the ground or condensed state is
also obtained exactly, and signals clearly the phase transition.
The application of these techniques to fermionic systems is also
briefly discussed.
\end{abstract}

\vspace{0.3in}
]

\section{Introduction}
\label{sec:Intro}

Recently there has been a renewed interest in the thermodynamics of
mesoscopic systems.  In particular, the recent experimental observation of
condensation phenomena in nanokelvin aggregates of atoms moving in magnetic
traps~\cite{Anderson-MH:1995science,Petrich-W:1995prl,Bradley-CC:1995prl,Davis-KB:1995prl}
has encouraged a number of recent theoretical investigations of this
phenomenon~\cite{Grossmann-S:1995pla,Kirsten-K:1996plb,Kirsten-K:1996pla}
extending earlier
investigations~\cite{Bagnato-V:1987pra,deGroot-SR:1950prsla}.
These works assume that the systems can be 
adequately described by the grand canonical ensemble in which both the
energy and particle number are allowed to fluctuate.  
The canonical ensemble is probably more representative of the
experimental conditions and has also been recently
investigated~\cite{Bronsens-F:1996ssc,Tempere-J:1997ssc}.

In this work, we analyze the thermodynamics of bosons in the the grand
canonical, canonical and microcanonical ensembles by applying particularly
effective techniques that two of us have used before in describing
nuclear systems 
statistically~\cite{Chase-KC:1995prl,Chase-KC:1994prc1,Mekjian-AZ:1990prl}.
This reproduces the recursive formula for the canonical
partition function from a combinatorial argument which 
proffers several advantages over the earlier purely functional
treatment~\cite{Bronsens-F:1996ssc}.
In particular, this interpretation enables a simple and exact
determination of the expected occupancies of the energy levels in any
of the ensembles.  This allows the usual thermodynamic functions to
be expressed exactly and succinctly, not just the partition function.
Additionally, the method is related to the permutational decomposition
of the liquid $^{4}$He partition function proposed by
Feynman~\cite{Feynman-RP:1953prev1,Feynman-RP:1972-stat-mech}.
Previously, we have used this correspondence to specify partition
functions for isotopic nuclear systems~\cite{Chase-KC:1994prc2}.  In
this paper, the correspondence enables as a new result an exact
determination of the microcanonical partition function.

\section{The Partition Function for Identical Bosons or Fermions}
\label{sec:pf}

In statistical mechanics the partition function entirely specifies the
thermodynamics, so our first priority is to determine a method of
computing the partition function in the various ensembles.
We begin by considering the grand canonical partition function, and
develop a combinatorial method of extracting the canonical
and microcanonical partition functions from the larger ensemble.
In the process we develop exact expressions for the expected
occupation of the energy levels in all three ensembles.

\subsection{Grand canonical partition function}

In the grand canonical ensemble, where the particle number and energy
are unconstrained, the partition function $Z$
for a system of bosons occupying energy levels $\varepsilon_{k}$ with
degeneracy $g_{k}$ is given by
\begin{equation}
\ln Z = - \sum_{j \ge 0} g_{j} 
  \ln \left( 1 - e^{(\mu-\varepsilon_{j})/k_{B}T} \right) \;,
  \label{eq:bose-gcpf}
\end{equation}
where $\mu$ is the chemical potential.  Let us think of this as a sum
either over canonical $Z_{n}$ or microcanonical $Z_{n,E}$
partition functions, i.e.
\begin{equation}
Z = \sum_{n} z^{n} Z_{n} = \sum_{n,E} z^{n} w^{E} Z_{n,E} \;,
    \label{eq:pf-relations}
\end{equation}
where the fugacity $z$ is given by
$z = \exp \{ \mu/k_{B} T\}$ and $w = \exp \{-1/k_{B}T \}$.
Expanding Eq.~(\ref{eq:bose-gcpf}) as a power series in $z, w$
yields 
\begin{eqnarray}
\ln Z
  & = & \sum_{j=0}^{\infty} g_{j} \sum_{k=1}^{\infty}
        {z^{k} \over k} w^{k \varepsilon_{j}}
        \nonumber \\
  & = & \sum_{k=1}^{\infty} z^{k} x_{k} 
    =   \sum_{k=1}^{\infty} \sum_{j=0}^{\infty} z^{k} 
        w^{k\varepsilon_{j}} x_{jk} \;,
        \label{eq:bose-lnZ}
\end{eqnarray}
where we have introduced $x_{jk} = g_{j}/k$ and 
$x_{k} = \sum_{j} w^{k \varepsilon_{j}} x_{jk}$ for reasons which will
become clear shortly.
If we exponentiate and Taylor expand Eq.~(\ref{eq:bose-lnZ}) 
we see that
\begin{eqnarray}
Z & = & \sum_{\pi_{01}=0}^{\infty} 
        {(z w^{\varepsilon_{0}} x_{01})^{\pi_{01}} \over \pi_{01}!}
        \sum_{\pi_{02}=0}^{\infty}
        {(z^{2} w^{2 \varepsilon_{0}} x_{02})^{\pi_{02}} \over \pi_{02}!}
        \times \cdots
        \nonumber \\
  & = & \sum_{\{ \pi_{jk} \}}
        \prod_{jk} {(z^{k} w^{k \varepsilon_{j}} x_{jk})^{\pi_{jk}}
        \over \pi_{jk}!} \;,
        \label{eq:bose-gcZprod}
\end{eqnarray}
where $\pi_{jk} \ge 0$.

Notice that the exponents of $w, z$ in Eq.~(\ref{eq:bose-gcZprod}) are
simple functions of $\{\pi_{jk}\}$.  
Since the exponent of $w$ in the grand canonical partition function is
the total energy $E$ in the particular microcanonical component, the
exponent of $w$, $\sum_{k} k \varepsilon_{j} \pi_{jk}$ should be equal
to the energy $E$ for that particular state $\{\pi_{jk}\}$.
This $E = \sum_{j} n_{j} \varepsilon_{j}$, where $n_{j}$ is the
number of particles in energy state $j$, so the state $\vec{\pi}$ is
related to $\vec{n}$ by
\begin{equation}
n_{j} = \sum_{k} k \pi_{jk} \;.
        \label{eq:nj-pijk}
\end{equation}
Combinatorially speaking, $\pi_{jk}$ counts collections of $k$
particles in the $j$th energy state.  In fact, 
$\{ \pi_{jk} : j {\rm\ fixed}\}$ is
simply a cycle class decomposition of a permutation of the $n_{j}$
particles in the $j$th energy level.  To see this, consider the
part of the grand canonical partition function due to energy level
$j$, namely the number of states of a system
with $n_{j}$ bosons in a level with degeneracy $g_{j}$, 
\begin{eqnarray}
{n_{j} + g_{j} - 1 \choose n_{j}}
  & = & \sum_{\{ \pi_{jk} : j {\rm\ fixed}\}}
        \prod_{k} {1 \over \pi_{jk}!} \left( {g_{j} \over k}
        \right)^{\pi_{jk}}
        \nonumber \\
  & = & {1 \over n_{j}!} \sum_{p \in S_{n_{j}}} 
        \prod_{k} g_{j}^{\pi_{jk}(p)} \;.
\end{eqnarray}
Here we see that 
${n_{j} + g_{j} - 1 \choose n_{j}}$ can be described in two ways.
The first is just the isolated contribution from the
above grand canonical partition function.  The second is a sum over
permutations of the $n_{j}$ particles, where $\pi_{jk}(p)$ is the
number of $k$-cycles in the permutation $p$ of $n_{j}$.  The
equivalence of these two expression is due to Cauchy and
Sylvester~\cite{Andrews-GE:1976-ToP}.

Feynman considered such permutations
in the condensation of liquid $^{4}$He~\cite{Feynman-RP:1972-stat-mech}
and was able to obtain a form for the partition function by expanding
the symmetrized density matrix as a sum over such permutations.
Specifically, the sum $\pi_{k} = \sum_{j} \pi_{jk}$,
the total number of cycles of length $k$, figures prominently in
Feynman's approach.
Such permutational decomposing of the partition function has also been
useful in our research on isotopic nuclear
systems~\cite{Chase-KC:1994prc2}, where the dual requirement of proton
and neutron conservation was considered as restricting the
permutational states to have been colored in a particular way.  Here
we can accomplish a similar interpretation by coloring the clusters
which form according to the energy levels they reside in.
Parallels of permutation problems with cluster yields were also noted
in~\cite{Chase-KC:1994prc1,Mekjian-AZ:1990prl}.

Since $\pi_{jk}$ has a combinatorial interpretation and is related to
$n_{j}$, it is important to obtain expectation values of it.
This is possible since each term of Eq.~(\ref{eq:bose-gcZprod}) can be
interpreted as an unnormalized probability for the state 
$\{ \pi_{jk} \}$, the normalization being simply the partition function.
Equation~(\ref{eq:bose-gcZprod}) then suggests we can evaluate such
expectation values of $\pi_{jk}$ by taking various derivatives of the
partition function.  More specifically, it is true that
$\langle \pi_{jk} \rangle = (x_{jk}/Z) (\partial Z/\partial x_{jk})$
and 
$\langle \pi_{jk} (\pi_{lm}-\delta_{jk,lm}) \rangle
  = (x_{jk} x_{lm}/Z) (\partial^{2} Z/\partial x_{jk} \partial x_{lm})$,
where $\langle \cdot \rangle$ denotes a grand canonical ensemble
average.  Now according to Eq.~(\ref{eq:bose-lnZ}), 
$Z = \exp \sum_{jk} z^{k} w^{k \varepsilon_{j}} x_{jk}$, so that
\begin{equation}
{\partial Z \over \partial x_{jk}} 
  = z^{k} w^{k \varepsilon_{j}} Z({\mathbf{x}}) \;.
    \label{eq:bose-Dgcpf}
\end{equation}
We then see that
\begin{eqnarray}
\langle \pi_{jk} \rangle 
  & = & z^{k} w^{k \varepsilon_{j}} x_{jk} \;,
        \\
\langle \pi_{jk} (\pi_{lm}-\delta_{jk,lm}) \rangle
  & = & z^{k+m} w^{k \varepsilon_{j}+ m \varepsilon_{l}} 
        x_{jk} x_{lm} \;.
\end{eqnarray}
We can now ``solve'' the grand canonical partition function now by using
the combinatorial interpretation to specify $w, z$, i.e.
by applying the constraints 
$n = \sum_{j} \langle n_{j} \rangle = \sum_{jk} k \langle \pi_{jk} \rangle$, 
$E = \sum_{j} \langle n_{j} \rangle \varepsilon_{j} = \sum_{jk} k
\langle \pi_{jk} \rangle \varepsilon_{j}$.

The analysis illustrated above for a system of
bosons applies equally well to a system of fermions.
In this case, the appropriate partition function is
\begin{equation}
\ln Z = \sum_{j} g_{j} \ln (1 + e^{(\mu-\varepsilon_{j})/k_{B}T}) \;.
\end{equation}
Expanding this yields Eq.~(\ref{eq:bose-lnZ}) with 
$x_{jk} = (-1)^{k+1} g_{j}/k$.  
The fact that $x_{jk}$ can be negative means that the combinatorial
interpretation is somewhat suspect, since some contributions to the
partition function are negative.  Nevertheless, the mathematics is
sound and expectation values of $\langle n_{j} \rangle$ can be
computed correctly by the same technique used above.
This fermionic case will be discussed in detail in another paper.

\subsection{Canonical partition function}

We can obtain the canonical partition function from the grand
canonical partition function by making restrictions on the sum
appearing in Eq.~(\ref{eq:bose-gcZprod}), i.e. if we restrict the sum
to terms where $n = \sum_{jk} k \pi_{jk}$ we will obtain the canonical
partition function.  
We could simplify some of the computations by defining 
$\pi_{k} = \sum_{j} \pi_{jk}$, so that the states are described by 
$\{ \pi_{k} \}$ where $\sum_{k} k \pi_{k} = n$.  Mathematically
speaking, such vectors $\{ \pi_{k} \}$ describe 
partitions of the integer $n$~\cite{Andrews-GE:1976-ToP}, where
$\pi_{k}$ is the number of time $k$ appears in such a partition.
However, sometimes it is more convenient to work instead with
$\pi_{jk}$.

Let us define $\pi(n) = \{ \pi_{jk} : \sum_{jk} k \pi_{jk} = n \}$.
Then the canonical partition function is given by
\begin{equation}
Z_{n}({\mathbf{x}}) = \sum_{\pi(n)}
  \prod_{k>0} {(w^{k \varepsilon_{j}} x_{jk})^{\pi_{jk}} \over \pi_{jk}!} \;.
  \label{eq:Zn}
\end{equation}
Since $\sum_{jk} k \pi_{jk} = n$ for each state in the ensemble, it is
true that $\sum_{jk} k \langle \pi_{jk} \rangle_{n} = n$, where 
$\langle \cdot \rangle_{n}$ denotes a canonical ensemble average.
The canonical ensemble average for $\pi_{jk}$ can be computed by
taking a derivative of the partition function
$\langle \pi_{jk} \rangle_{n} 
  = (x_{jk}/Z_{n}) (\partial Z_{n}/\partial x_{jk})$
as can be seen from Eq.~(\ref{eq:Zn}).
In terms of the canonical partition functions, the canonical
constraint can be rewritten then as
\begin{equation}
n Z_{n}({\mathbf{x}})
  = \sum_{k=1}^{n} k \sum_{j \ge 0} x_{jk}
          {\partial Z_{n} \over \partial x_{jk}} \;.
          \label{eq:Zn-recursion}
\end{equation}
If we could relate $\partial Z_{n}/\partial x_{jk}$ to 
$Z_{1}({\mathbf{x}}), \ldots, Z_{n-1}({\mathbf{x}})$ then the above
identity would imply a recursive method for computing
$Z_{n}({\mathbf{x}})$.

In fact $\partial Z_{n}/\partial x_{jk}$ is precisely
proportional to a particular partition function as can be seen by 
combining Eq.~(\ref{eq:bose-Dgcpf}) and Eq.~(\ref{eq:pf-relations}),
yielding 
\begin{equation}
{\partial Z_{n} \over \partial x_{jk}} 
  = w^{k \varepsilon_{j}} Z_{n-k}({\mathbf{x}}) \;.
\end{equation}
So the identity given in Eq.~(\ref{eq:Zn-recursion}) is in fact a
recursion for construction the partition function:
\begin{equation}
n Z_{n}({\mathbf{x}}) 
  = \sum_{jk} k x_{jk} w^{k \varepsilon_{j}} Z_{n-k}({\mathbf{x}})
  = \sum_{k} k x_{k} Z_{n-k}({\mathbf{x}}) \;,
\end{equation}
with $Z_{0}({\mathbf{x}}) = 1$ as the anchor for the recursion.
The expectation values for $\pi_{jk}$ are simply related to the
partition functions:
\begin{eqnarray}
\langle \pi_{jk} \rangle_{n}
  & = & w^{k \varepsilon_{j}} x_{jk} 
        {Z_{n-k}({\mathbf{x}}) \over Z_{n}({\mathbf{x}})} \;,
        \label{eq:pijk_n}
        \\
\langle \pi_{jk} (\pi_{lm}&-&\delta_{jk,lm}) \rangle_{n}
        \nonumber \\
   & = & w^{k \varepsilon_{j}+ m \varepsilon_{l}} 
        x_{jk} x_{lm}
        {Z_{n-k-m}({\mathbf{x}}) \over Z_{n}({\mathbf{x}})} \;.
        \label{eq:pijklm_n}
\end{eqnarray}

\subsection{Microcanonical partition function}

A similar combinatorial argument can extract the microcanonical partition
function from the canonical or grand canonical partition function.
Let us define $\pi(n,E) = \{ \pi_{jk} : \sum_{jk} k \pi_{jk} = n, 
\sum_{jk} k \varepsilon_{j} \pi_{jk} = E\}$.  Then
\begin{eqnarray}
Z_{n,E}({\mathbf{x}})
  & = & \sum_{\pi(n,E)}
        \prod_{jk} {x_{jk}^{\pi_{jk}} \over \pi_{jk}!} \;.
\end{eqnarray}
The partition function can now be developed by recursion by noting that
\begin{eqnarray}
\langle \pi_{jk} \rangle_{n,E}
  & = & x_{jk} {Z_{n-k,E-k\varepsilon_{j}}({\mathbf{x}}) \over 
        Z_{n,E}({\mathbf{x}})} \;,
        \label{eq:pijk-mc}
        \\
\langle \pi_{jk} (\pi_{lm}&-&\delta_{jk,lm}) \rangle_{n,E}
        \nonumber \\
  & = & x_{jk} x_{lm}
        {Z_{n-k-m,E-k\varepsilon_{j}-m\varepsilon_{l}}({\mathbf{x}}) 
        \over Z_{n}({\mathbf{x}})} \;,
\end{eqnarray}
and applying the identity 
$\sum_{jk} k \langle \pi_{jk} \rangle_{n,E} = n$, yielding
\begin{equation}
n Z_{n,E}({\mathbf{x}}) =
  \sum_{jk} k x_{jk} Z_{n-k,E-k\varepsilon_{j}}({\mathbf{x}}) \;,
\end{equation}
with $Z_{0,E}({\mathbf{x}}) = \delta_{0,E}$.

A similar situation arises in models of nuclear fragmentation, where
there is also two constraints, namely the conservation of proton and
neutron number, in which case~\cite{Chase-KC:1994prc2}
it is useful to restrict the ensemble by summing over
isotopes or isobars.  Then the existence of permutational
Cauchy identities simplified the resulting partition functions.  
In this case the same kind of mechanism also applies.  We can sum over
the energy states and arrive at the $\pi_{k}$ cluster representation
of the states.  Or we can sum over the cluster sizes and arrive at the
$n_{j}$ occupation number representation.  In each case, the weight
retains its form and the analysis in the restricted space is identical
or nearly identical as in the large space.

\section{Application to the Harmonic Oscillator}
\label{sec:harmonic}

Having the partition function now allows us to compute the
thermodynamics in the traditional manner.
For the case of particles moving
in a $d$ dimensional isotropic harmonic potential, $g_{j}$ and
$\varepsilon_{j}$ can be
evaluated directly.
The Hamiltonian for the system is given by
$\hat{H} = \sum_{i=1}^{d} p_{i}^{2}/2m
   + m \omega^{2} \sum_{i=1}^{d} q_{i}^{2}/2$
where $(q_{i}, p_{i})$ is the position and momentum in the $i$th
direction.  So $H = \sum_{i} H_{i}$ where $H_{i}$ is the Hamiltonian
for a one-dimensional oscillator acting on $(q_{i}, p_{i})$.
Of course 
$H_{i} \psi_{n}(q_{i}) = (n+1/2) \hbar \omega \psi_{n}(q_{i})$, and by 
separation of variables
$H \psi_{n_{1}, \ldots, n_{d}} = (n_{1} + \cdots + n_{d} + d/2) \hbar \omega \psi_{n_{1}, \ldots, n_{d}}$.  
If we write this as simply $\varepsilon_{j} = (j+d/2)\hbar\omega$, we
see that the $j$th energy level has a degeneracy $g_{j}$ based on the
number of ways one can write $j$ as an ordered sum of $d$ nonnegative
integers, i.e. $j=n_{1}+\ldots+n_{d}$.
In combinatorics this is known as a $d$-composition of $j$,
and there are $g_{j} = {j+d-1 \choose j}$ ways of doing this.
Substituting this into $x_{k} = (1/k) \sum_{j} g_{j} w^{k \varepsilon_{j}}$
yields 
\begin{eqnarray}
x_{k}
  & = & {1 \over k} \sum_{j=0}^{\infty}
        {j+d-1 \choose j} x^{k(j+d/2)}
    =   {1 \over k} {x^{k d/2} \over (1 - x^{k})^{d}} \;,
        \label{eq:x_k-HO}
\end{eqnarray}
where $x = \exp \{ -\hbar\omega/k_{B}T \}$ and we have applied
the negative binomial identity 
$(1-x)^{-d} = \sum_{n \ge 0} x^{n} {n+d-1 \choose n}$.

\subsection{Canonical thermodynamics}

We can now compute the canonical partition function for the harmonic
oscillator recursively using Eq.~(\ref{eq:Zn-recursion}).
Of interest is the fact that in one dimension the
recursion yields a particularly simple form due to a partition theorem of
Cayley (cf.~\cite{Andrews-GE:1976-ToP}, pg.~209), namely:
$Z_{n,d=1}(x) 
  = x^{n/2}/((1-x)(1-x^{2}) \cdots (1-x^{n}))$.
This partition function was investigated in a different context by
A.~Z. Mekjian and S.~J. Lee~\cite{Mekjian-AZ:1991pra1}.
In two and higher dimensions, the partition function does not have a
simple closed form, but can be computed by the above recursion.

Having obtained the partition function we can compute any
thermodynamic function of interest by a judicious use of partial
derivatives.   For example, in the canonical ensemble,
the internal energy
$U = k_{B} T^{2} ({\partial/\partial T})_{V} \ln Z_{n}$
and the specific heat $C_{V} = (\partial U/\partial T)_{V}$ for a
group of bosons are given by
\begin{eqnarray}
{U \over k_{B}T} 
  & = & \sum_{k>0} {T \over x_{k}} 
        {\partial x_{k} \over \partial T} \langle \pi_{k} \rangle_{n} \;,
  \label{eq:internal-energy}
        \\
{C_{V} \over k_{B}} 
  & = & {2 U \over T} + \left[ \sum_{k>0} \left( 
        {T^{2} \over x_{k}} {\partial^{2} x_{k} \over \partial T^{2}}
        -\left({T \over x_{k}} {\partial x_{k} \over \partial T} \right)^{2}
        \right) \langle \pi_{k} \rangle_{n} \right.
        \nonumber \\
      &&\hspace{-0.3in}
      + \sum_{jk} \left. {T \over x_{j}} {\partial x_{j} \over \partial T}
        {T \over x_{k}} {\partial x_{k} \over \partial T}
        \left( \langle \pi_{j} \pi_{k} \rangle_{n} - 
        \langle \pi_{j} \rangle_{n} \langle \pi_{k} \rangle_{n} 
        \right) \right] \;.
   \label{eq:cv}
\end{eqnarray}
The pressure 
$P = k_{B}T ({\partial/\partial V})_{T} \ln Z_{n}$
and the incompressibility 
${1/\kappa_{T}} = -V ({\partial P/\partial V})_{T}$ can be computed in
a similar fashion.
For the harmonic oscillator, the partial derivatives of $x_{k}$ with
respect to $T$ are readily determined.  The partial derivatives with
respect to $V$ can also be determined, once an appropriate volume is specified.
Since $\hbar\omega$ determines the size of the trap, $\hbar\omega$
determines the volume.  Assuming $h^{2}/2m\lambda^{2} \approx m \omega^{2}
\lambda^{2}/2$ where $\lambda$ is a particle's wavelength we see that
$\hbar\omega \propto V^{-2/d}$.

For computational purposes, the above expressions are concise and
powerful, but for understanding the source of the phase transition,
one needs to rewrite the above expressions in terms of the occupancies
of the levels.  Since 
$Z_{n} = \sum_{\vec{n}} Z_{\vec{n}} \exp \{-\sum_{j} n_{j}/k_{B}T \}$
it can be shown that
$U = \sum_{j} \langle n_{j} \rangle_{n} \varepsilon_{j}$, and that
$C_{V} = \sum_{jk} \left( \langle n_{j} n_{k} \rangle_{n} -
            \langle n_{j} \rangle_{n} \langle n_{k} \rangle_{n} \right)
            \varepsilon_{j} \varepsilon_{k}/(k_{B}T^{2})$.
As the critical point is marked by a peak in the specific heat, from
the above expression its source must be due to large fluctuations in
the occupation numbers.  The expected occupation numbers can be
inferred from Eq.~(\ref{eq:nj-pijk}) and
Eqs.~(\ref{eq:pijk_n})~and~(\ref{eq:pijklm_n}), yielding 
\begin{eqnarray}
\langle n_{j} \rangle_{n}
  & = & g_{j} \sum_{k} w^{k\varepsilon_{j}}
        {Z_{n-k}({\mathbf{x}}) \over Z_{n}({\mathbf{x}})} \;,
        \\
\langle n_{j} n_{k} \rangle_{n}
  & = & \delta_{jk} g_{j}  \sum_{m} m w^{m \varepsilon_{j}}
        {Z_{n-m}({\mathbf{x}}) \over Z_{n}({\mathbf{x}})}
        \nonumber \\
        &&+ g_{j} g_{k} \sum_{lm} w^{l \varepsilon_{j} + m \varepsilon_{k}} 
        {Z_{n-l-m}({\mathbf{x}}) \over Z_{n}({\mathbf{x}})} \;,
\end{eqnarray}
and allow the expected occupation and its fluctuation to be computed
readily.

\subsection{Microcanonical thermodynamics}

In the microcanonical ensemble, the temperature is defined by 
$T = (\partial E/\partial S)_{n,V}$, and this allows us to compute the
thermodynamic functions, e.g. for the harmonic oscillator
\begin{eqnarray}
C_{V} 
  & = & - {(S_{n,E+\hbar\omega}-S_{n,E})(S_{n,E}-S_{n,E-\hbar\omega}) \over
           S_{n,E-\hbar\omega}-2 S_{n,E} + S_{n,E+\hbar\omega}} \;,
\end{eqnarray}
where $S_{n,E} = \log Z_{n,E}$.
Relating the thermodynamics to the occupancies of the levels is more
difficult than in the canonical case, since the temperature is not an
independent parameter.

\section{Discussion and Conclusion}
\label{sec:discussion}

\begin{figure}
\caption{The specific heat and ground state occupation in three
dimensions in the grand canonical, canonical and microcanonical
ensembles for a system of 100 particles.  The curves become nearly
identical for larger groups of particles.}
\label{fig:cv-gs}
\centerline{\includegraphics[bb = 122 11 535 764, angle = -90,
            width = 7.0in]{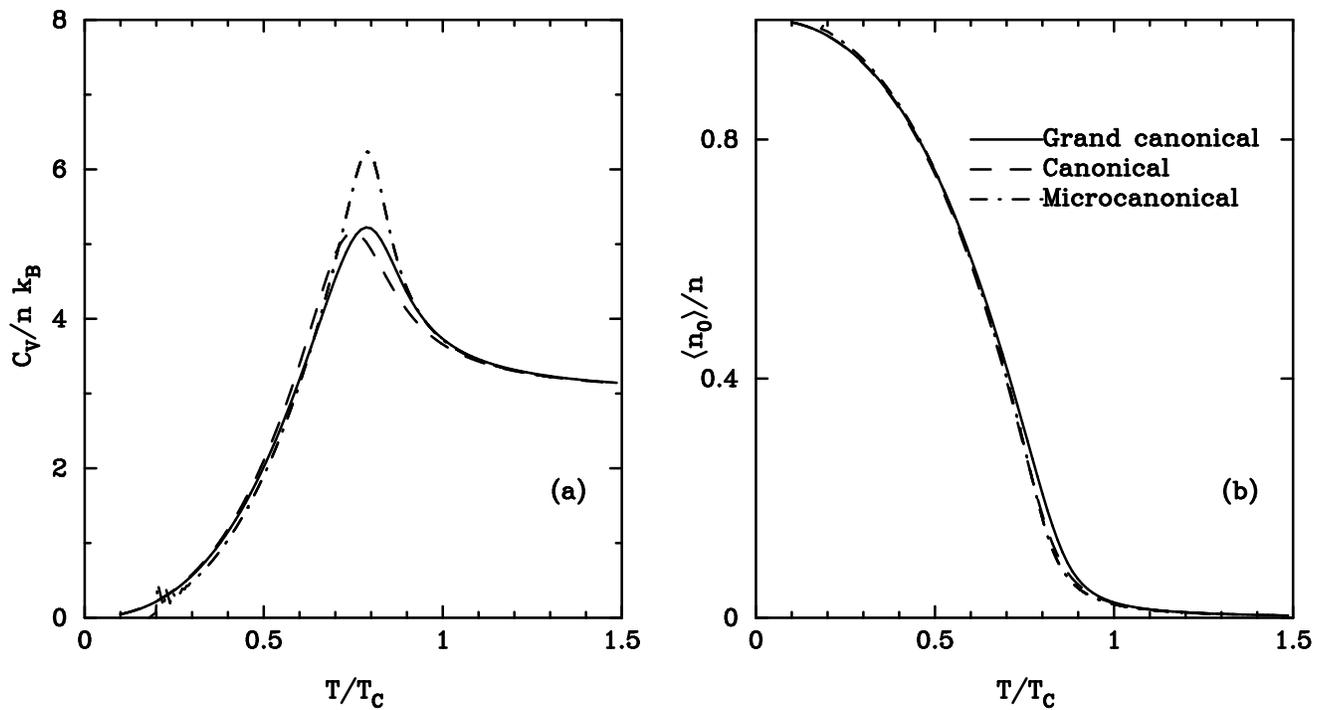}}
\end{figure}

As an application of the above techniques, we display the thermodynamics
of particles in a harmonic trap in the three ensembles.  The 
specific heat (Fig.~\ref{fig:cv-gs}(a))
and the occupation of the ground state (Fig.~\ref{fig:cv-gs}(b)) are
substantially in agreement in all 
three ensembles, confirming the essential validity of the use of the
different ensembles even for such small groups of particles.  The
finite size effects are however measurable, and should not be
completely discounted, especially in smaller systems.

In summary, we have introduced a combinatorial approach to computing
thermodynamic functions in systems of identical bosons.  It is
ideally suited to systems in which the energy levels are separated by
integral multiples of some fundamental energy, such as in the harmonic
oscillator ($\varepsilon_{j} = j\hbar\omega$) and the rigid rotator
($\varepsilon_{j} = j(j+1)\hbar^{2}/2I$).  Such techniques can also be
applied to groups of identical fermions, which has applications to
the nuclear shell model at finite temperatures.

\acknowledgements

This work was supported by Department of Energy Grants
DE-FG02-95ER40940 and DE-FG02-96ER40987.


\hsize\textwidth\columnwidth\hsize\csname @twocolumnfalse\endcsname
\widetext

\end{document}